\documentclass[graybox]{svmult}

\usepackage{type1cm}        % activate if the above 3 fonts are
\usepackage{makeidx}         % allows index generation
\usepackage{graphicx}        % standard LaTeX graphics tool
\usepackage{multicol}        % used for the two-column index
\usepackage[bottom]{footmisc}% places footnotes at page bottom
\usepackage{booktabs}
\usepackage{natbib}
\usepackage{framed,color}
\definecolor{shadecolor}{rgb}{.7,.7,.7} % change to suitable shade of grey

\usepackage{newtxtext}       % 
\usepackage{newtxmath}       % selects Times Roman as basic font

\makeindex             % used for the subject index
\begin{document}

\title*{Challenges in Survey Research}
\author{Stefan Wagner, Daniel Mendez, Michael Felderer, Daniel Graziotin and Marcos Kalinowski}
\authorrunning{Stefan Wagner et al.}
\institute{Stefan Wagner \at University of Stuttgart, Stuttgart, Germany, \email{stefan.wagner@iste.uni-stuttgart.de}
\and Daniel Mendez \at Technical University of Munich, Munich, Germany\\ Blekinge Institute of Technology, Karlskrona, Sweden \\ fortiss GmbH, Munich, Germany \\
ORCID: 0000-0003-0619-6027 
\email{mendezfe@acm.org}
\and Michael Felderer \at University of Innsbruck, Innsbruck, Austria \\
Blekinge Institute of Technology, Karlskrona, Sweden \\ 
\email{michael.felderer@uibk.ac.at}
\and Daniel Graziotin \at University of Stuttgart, Stuttgart, Germany, \email{daniel.graziotin@iste.uni-stuttgart.de}
\and Marcos Kalinowski \at Pontifical Catholic University of Rio de Janeiro, Rio de Janeiro, Brazil, \email{kalinowski@inf.puc-rio.br}}
\maketitle

\abstract{While being an important and often used research method, survey research has been less often discussed on a methodological level in empirical software engineering than other types of research. This chapter compiles a set of important and challenging issues in survey research based on experiences with several large-scale international surveys. The chapter covers theory building, sampling, invitation and follow-up, statistical as well as qualitative analysis of survey data and the usage of psychometrics in software engineering surveys.}

\section{Introduction}
\label{sec:intro}

Empirical software engineering started with a strong focus on controlled experiments. It widened only later to case studies and similar research methods. Both methodologies have been discussed extensively for software engineering~\citep{wohlin2012,runeson2012}. While survey research has been used to capture a broader sample for mostly cross-sectional studies, the methodological issues have rarely been discussed. 

The aim of this chapter is to complement existing more general literature on survey research and questionnaire design as well as the existing software-engineering-specific literature. Therefore, this is not a tutorial to survey research, but it provides a compact description of important issues and lessons learned that any empirical software engineering research can make use of in their next surveys.

To not only discuss pure methodology and theory, we provide concrete examples of our experiences with the methodologies based on two lines of survey research: First, the project \emph{Naming the Pain in requirements engineering}\footnote{\url{http://napire.org}} (NaPiRE) has the goal to provide an empirical basis for requirements engineering research by capturing the state of the practice and current problems and challenges with requirements engineering. We have already made three rounds of surveys in this project over seven years and over ten countries. In these, we developed a theory as basis for the questionnaire, several variations on questions for similar concepts and also experimented with different methodological options~\citep{MENDEZFERNANDEZ2015616,Wagner:2019:SQR:3316413.3306607,fernandez2017naming}. We will discuss these variations and experiences in the following.

We complement the NaPiRE experiences with a study that aimed to assess the happiness of software developers and targeted GitHub developers with a pyschometrically validated test~\citep{graziotin2019,graziotin2018happens,graziotin2017unhappy}. This example, described in Section~\ref{ssec:psych:example} is different in the target population and how the questionnaire was created. Hence, it gives us even more possibilities to discuss.

The chapter is organized so that we discuss different areas that we consider interesting and challenging. We start with a discussion on how survey research can be integrated with theory building, then explain what we need to consider when using psychometric tests in our questionnaires and why we need to consider psychometric properties. We then discuss the limited possibilities in evaluating the sample of a survey study including a short discussion of ethics. We continue with the closely related issue of how and whom to invite to a survey and how to manage follow-ups. The last two sections discuss issues in quantitative statistical and qualitative analysis of the data from a survey.

\section{Survey Research and Theory Building}
\label{sec:survey-design}

The ultimate goal of empirical software engineering is, in one way or another, to build and evaluate scientific theories by applying empirical research methods~\citep{MendezPassoth18}. Survey research is one such means to contribute to theory development~\citep{malhotra1998assessment} as the main objective for conducting a survey is either of the following~\citep{wohlin2012,pinsonneault1993survey}: \emph{explorative}, \emph{descriptive} or \emph{explanatory}. Explorative surveys are used as a pre-study to a more thorough investigation to assure that important issues like constructs in a theory like requirements elicitation techniques are not foreseen. Descriptive surveys can be conducted to enable assertions about some population like the distribution of certain attributes (e.g., usage of requirements elicitation techniques). The concern is not why the observed distribution exists, but instead what that distribution is. Finally, explanatory surveys aim at making explanatory claims about the population (e.g., why specific requirements elicitation techniques are used in specific contexts). 

A theory provides explanations and understanding in terms of basic constructs and underlying mechanisms, which constitute an important counterpart to knowledge of passing trends and their manifestation~\citep{hannay2007systematic}. The main aim of a theory is to \emph{describe}, \emph{explain} or even \emph{predict} phenomena, depending on the purpose of the theory~\citep{gregor2006nature}. A theory can be defined as a statement of relationship between units observed or approximated in the empirical world~\citep{malhotra1998assessment}, i.e. we capture a pattern in real world phenomena. Theories may have further quality criteria, such as the level of support or practical and/or a scientific utility~\citep{stol2015theory}. 

From the practical perspective, theories should be useful and explain or predict phenomena that occur in software engineering. From a scientific perspective, theories should guide and support further research in software engineering. The main building blocks of a theory according to \citet{sjoberg2008building} are constructs, relationships, explanations, and a scope. Constructs describe what the basic elements are, propositions how the constructs interact with each other, explanations why the propositions are as specified, and the scope elaborates what the universe of discourse is in which the theory is applicable.  

The five steps of theory building of \citet{sjoberg2008building} are 
\begin{enumerate}
    \item defining the constructs,
    \item defining the propositions,
    \item providing explanations to justify the theory,
    \item determining the scope, and
    \item testing the theory (or, more precisely, to test its consequences via hypotheses, i.e. testable propositions) through empirical research
\end{enumerate}

For the last steps, mainly controlled experimentation are typically considered. In general, the relationship of theory building and experiments has been well investigated in software engineering~\citep{hannay2007systematic}, whereas the relationship to survey research has not. Theory building and evaluation can guide the design and analysis of surveys, and surveys can also be applied to test theories. In the following, we discuss the interplay between theory building and survey research based on examples taken from NaPiRE. 

In the early stages of studying a phenomenon, concepts of interest need to be explored and described in a conceptual framework or theory defining basic constructs and relationships, which also corresponds to the initial steps of theory building, i.e. definition of constructs and propositions. In later phases, a phenomenon can be explained and finally predictions based on cause-effect relationships can be drawn. Survey research can support all these phases of theory building. Both activities, survey research and theory-building, are strongly interrelated. The concrete relationship between survey research and theory-building depends on whether the theory is descriptive, explanatory, or predictive.

Initial theories, that is to say theories for which the level of evidence is yet weak, can be drawn from observations and available literature. An initial theory can be a taxonomy of constructs~\citep{usman2017taxonomies} or a set of statements relating constructs. \citet{inayat2015systematic} provide, for instance, an initial taxonomy on practices adopted in agile RE according to published empirical studies. Also common terminology as, for instance, provided by the Software Engineering Body of Knowledge (SWEBoK) by \citet{bourque2014guide}, which covers a taxonomy of requirements elicitation techniques, can be considered as an initial descriptive theory. For NaPiRE, we followed a similar strategy where we elaborated a set of constructs and propositions based on available literature and expert knowledge, thus, unifying isolated studies to a more holistic but initial (descriptive) theory on contemporary practices in RE~\citep{MENDEZFERNANDEZ2015616}. One such example was how practitioners tend to elicit requirements. Exemplary statements for the requirements elicitation were \emph{Requirements are elicited via workshops} or \emph{Requirements are elicited via change requests}. These two statements relate the concept requirements elicitation to the concepts workshops and change requests, respectively. The survey is designed to test the theory and find next statements to extend the theory. The statements on requirements elicitation resulted  in the closed multiple-choice survey question "If you elicit requirements in your regular projects, how do you elicit them?" with the additional option "Other". The responses were that 80\% use workshops and discussions with the stakeholders, 58\% use change requests, 44\% use prototyping, 48\% refer to agile approaches at the customer's site, and 7\% use other approaches. The two statements from the theory about requirements elicitation via workshops and change requests, respectively, were supported by respective null hypothesis tests (see Sec.~\ref{sec:hypothesis_testing}).

Subsequent survey runs were then designed to test that initial theory and make further observations to further extend, refine and improve the initial theory to an explanatory theory.

In principle, the more advanced theories are, the better explanations for the propositions they provide. The core issue of this is to provide explicit assumptions and logical justifications for the constructs and propositions of the theory. Tab.~\ref{tab:e-elicitation} shows propositions and explanations for requirements elicitation as formulated in the theory presented by \citet{Wagner:2019:SQR:3316413.3306607}. The presentation follows the tabular schema for presenting explanatory theories as suggested by \citet{sjoberg2008building}.

\begin{table}[htb]
\centering 
\caption{Propositions about elicitation with explanations after the survey~\cite{Wagner:2019:SQR:3316413.3306607}\label{tab:e-elicitation}}{%
\begin{tabular}{lp{0.7\linewidth}l}
\toprule
\textbf{No.} & \textbf{Propositions}  \\\midrule
P~1 & Requirements are elicited via interviews.  \\
P~2 & Requirements are elicited via scenarios.  \\
P~3 & Requirements are elicited via prototyping.  \\
P~4 & Requirements are elicited via facilitated meetings (including workshops).  \\
P~5 & Requirements are elicited via observation.  \\ \midrule
\textbf{No.} & \textbf{Explanations} & \textbf{Propositions}  \\\midrule
E~1 & Interviews, scenarios, prototyping, facilitated meetings and observations allow the requirements engineers to include many different viewpoints including those from non-technical stakeholders & P1--P5 \\ 
E~2 & Prototypes and scenarios promote a shared understanding of the requirements among stakeholders & P2, P3\\\bottomrule
\end{tabular}}
\end{table}

\begin{figure}[htb]
    \centering
    \includegraphics[width=\textwidth]{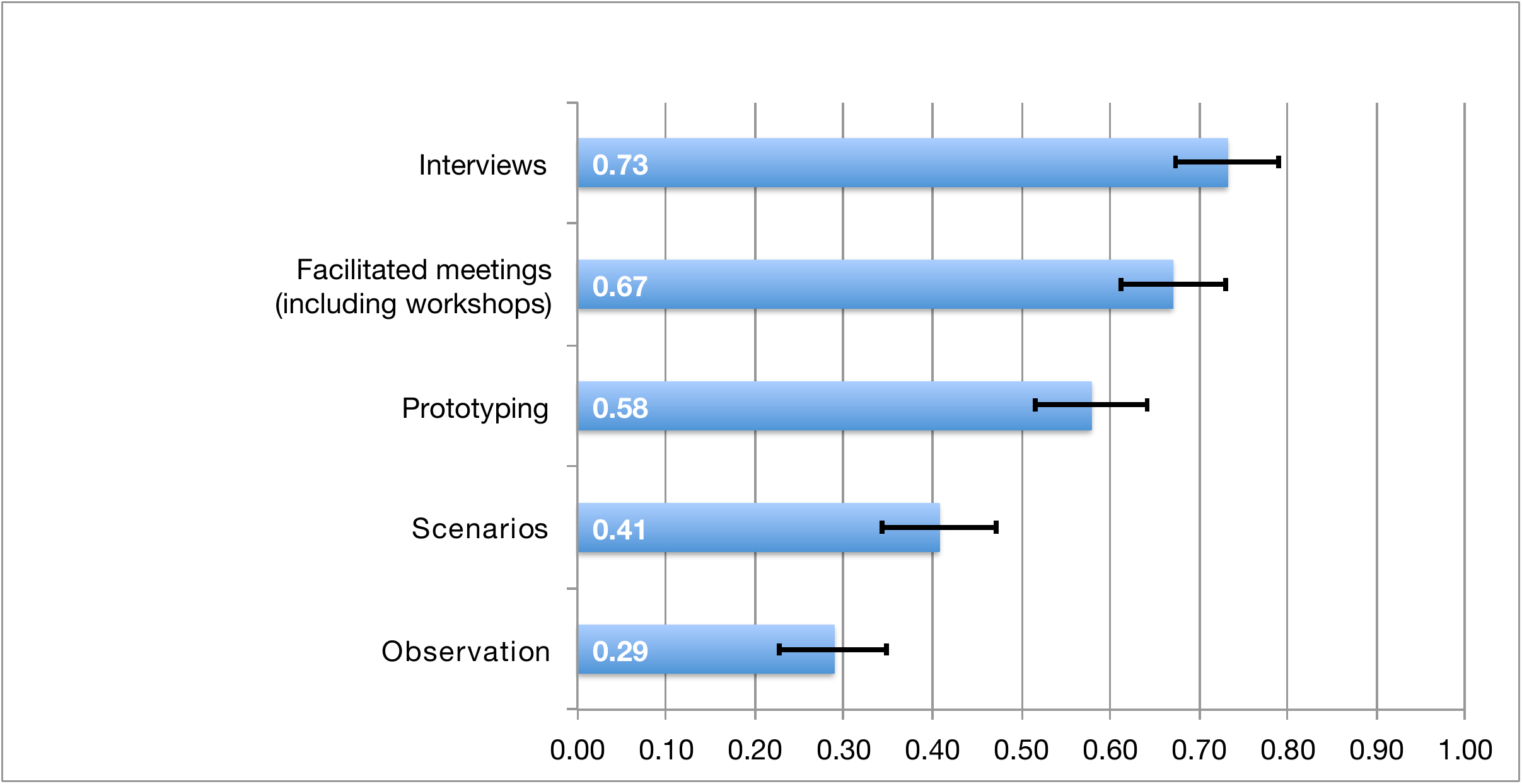}
    \caption{Proportions with confidence intervals for the question ``How do you elicit requirements?'' \cite{Wagner:2019:SQR:3316413.3306607}}
    \label{fig:confidence-intervals}
\end{figure}

The first run, however, showed that other elicitation techniques are also widely in use. This resulted in the propositions stated in Tab.~\ref{tab:e-elicitation}. P~1, P~2, P~3, and P~5 are new. P~4 was already supported in the first run and included in the initial theory. The used terminology in the propositions was also aligned with elicitation techniques as described in the SWEBoK. The answer possibilities in the questionnaire correspond directly to the propositions and resulted in the closed multiple-choice survey question "If you elicit requirements in your regular projects, how do you elicit them?" with the choices "Interviews", "Scenarios", "Prototyping", "Facilitated meetings (including workshops)", "Observation", and "Other". The answers of the respondents together with an error bar that represents the 95\% confidence interval (CI) is shown in Fig.~\ref{fig:confidence-intervals}. The confidence intervals of all response types, even from the least frequently used elicitation technique  Observations with $P = 0.29$ [0.23, 0.35] is still larger than the threshold of 0.2. We therefore have support for all corresponding propositions P~1 to P~5. Additional answers for ``others'' included ``Created personas and presented them to our stakeholders'', ``Questionnaires''/``Surveys'' , ``Analysis of existing system'' and ``It depends on the client.'' Especially some kind of surveys/questionnaires are mentioned several times. This could be a candidate for an additional proposition for the refined theory of the next iteration. 

The explanations E~1 and E~2 are for the five propositions P~1 to P~5. The difference between a proposition and an explanation is that the former is a relationship among constructs, and the latter is a relationship among constructs and other categories, which are not central enough to become constructs~\citep{sjoberg2008building}. For explanations of the propositions, we do not have any additional insights from the open answers, which would be of value. However, explanations can be backed up by literature. For instance, \citet{sommerville1998viewpoints} state that it is important to include different viewpoints during requirements elicitation, which supports E~1. \citet{mannio2001reqelicit} developed an iterative requirements elicitation method combining prototypes and scenarios, which supports E~2.

Predictive theories are geared towards predicting what will happen. The key underlying principle is finding cause-effect relationships among variables. This is typically performed via quantitative statistical models like correlation and regression. Correlation quantifies the degree to which variables are related. Regression quantifies the relationship between independent and dependent variables. In any case, causal quantitative relationships should always be backed up by theory-based expectations on how and why variables should be related. \citet{gregor2006nature} even considers integrated explanatory and predictive theories as a separate type of theory. Therefore, qualitative methods also play an important role even in predictive theories.

Based on the survey results of the second NaPiRE run, we developed cause-effect relationships with different degrees of quantification to support predictive theories.

\begin{figure}[htb]
    \centering
    \includegraphics[width=\textwidth]{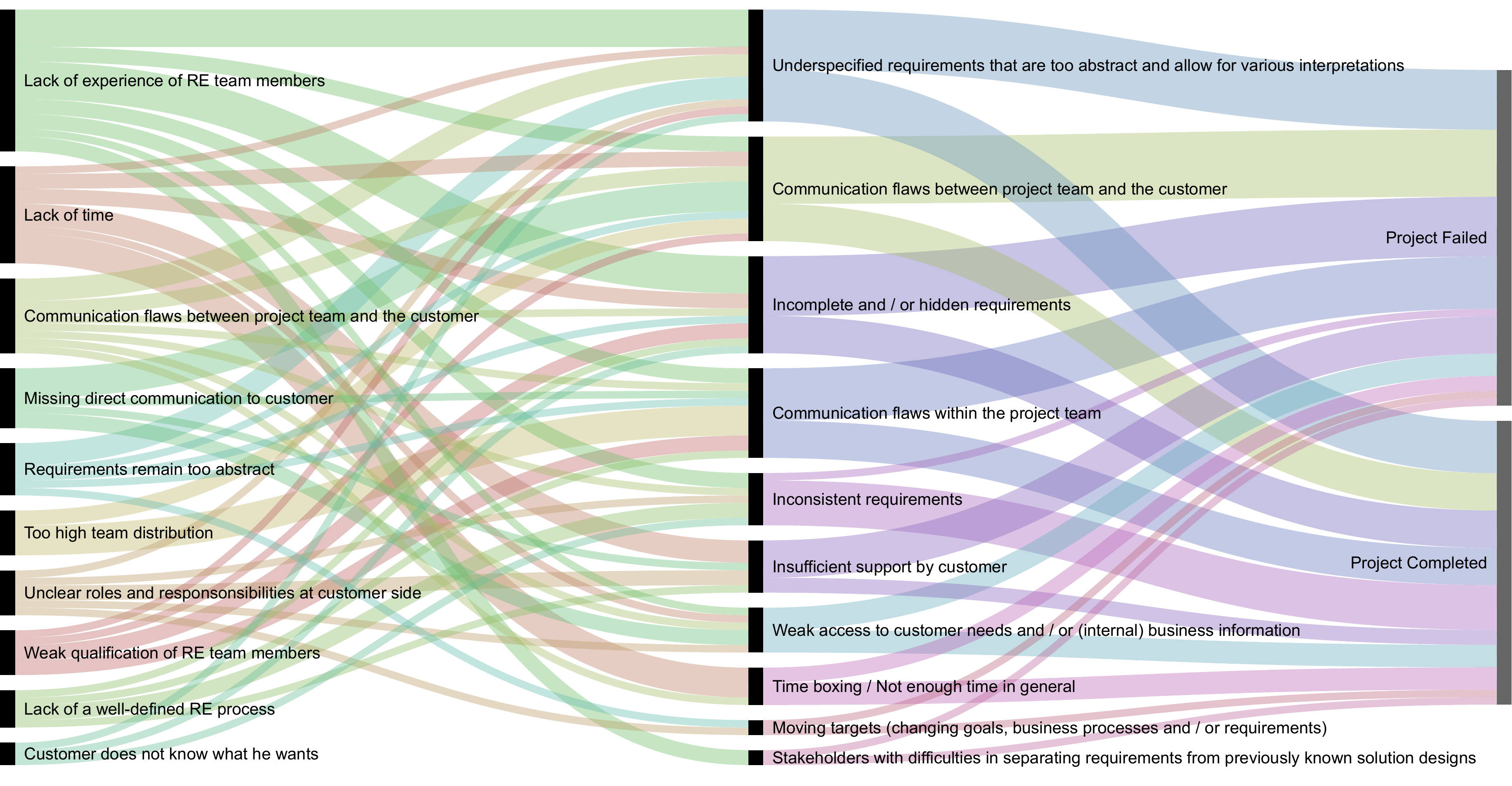}
    \caption{Cause-effect relationship between top ten causes, top ten RE problems and effects in terms of project impact \cite{fernandez2017naming}}
    \label{fig:alluvial-cause-effect}
\end{figure}

\citet{fernandez2017naming} developed initial cause-effect relationships between top ten causes, top ten RE problems as well as the project impact, i.e., whether projects failed or were completed. The resulting relationships are shown via an Alluvial diagram in Fig.~\ref{fig:alluvial-cause-effect}. This diagram enables initial predictions of the project impact based on available causes like lack of time or missing direct communication to customer and problems like incomplete and/or hidden requirements. 

The predictive model was created based on the survey in the following way. After selecting the five most critical RE problems in the survey, we asked our respondents to provide what they believe to be the main causes and effects for each of the problems. They provided the causes and effects in an open question format, with one open question for the cause and another for the effect for each the previously selected RE problems. For analyzing the answers given to the open questions on what causes and effects the RE problems have, we applied qualitative data analysis techniques as recommended in context of Grounded Theory (see Section~\ref{sec:qualitative-analysis}). For each answer given by a participant, we applied open and axial coding until reaching a saturation in the codes and relationships, and we allocate the codes to previously defined subcategories, i.e., Customer, Design/Implementation, Product, Project, and Verification/Validation for the effect. The procedure finally delivers triples of causes, problems, and effects that are visualized in the Alluvial diagram shown in Fig.~\ref{fig:alluvial-cause-effect}. Dynamically implemented Alluvial diagrams, which support highlighting and hiding elements, provide additional support for predictive analysis.

\citet{fernandez2018evidence} implemented the available cause-effect relationships in dependency of two context factors such as company size or software process model used as Bayesian networks (see Section~\ref{sec:bayesian_analysis} for Bayesian analysis). We did so based on similar work done also for defect causal analyses~\citep{kalinowski2017supporting}. Our implementation allowed us to quantify the conditional probabilistic distributions of all phenomena involved. More precisely, it allows us to, based on the NaPiRE data used as learning set, use the Bayesian network inference to obtain the posterior probabilities of certain phenomena to occur when specific causes are known. This supports evidence-based risk management in requirements engineering. 

So in the Bayesian approach, we performed a cross-sectional analysis of the NaPiRE data from one run (i.e., the second run), by blocking the data based on specific information from the survey like company size or the software development process used. Blocking can also help to refine the scope of a theory. In future, if NaPiRE will have been replicated several times, even a longitudinal analysis will be possible to develop predictive theories by analyzing time series.

\begin{shaded}\textbf{Take Aways}

\begin{itemize}
    \item Survey research and theory-building are strongly interrelated. The exact relationship depends on whether the theory is descriptive, explanatory, or predictive.
    \item Survey data supports the definition or refinement of constructs, relationships, explanations, and the scope of a theory as well as testing of a theory. 
    \item Theories are of high value to guide the design of surveys.
\end{itemize}

\end{shaded}

\section{Issues in Sampling}
\label{sec:sampling}

At the beginning of any design of survey research, we should clarify what the target population is that we try to characterize and generalize to. Statistical analysis (see Sec.~\ref{sec:statistical}), which we apply to survey data, relies on systematic sampling from this target population. In software engineering surveys, the unit of analysis that defines the granularity of the target population are often \citep{deMello2015} 
\begin{itemize}
    \item an organization, 
    \item a software team or project, or
    \item an individual.
\end{itemize}

For common research questions, we are typically interested in producing results related to all organizations that develop software in the world or all software developers in the world. Sometimes, this is reduced to certain regions of the world such as all requirements engineers in Europe or Brazil. The reason for this large aim in the target population is that we want to find theories that have a scope as wide as possible.

This brings us, however, to the problem that in most cases, we have no solid understanding about the target population. How many software developers are there in the world? Which companies are developing software? What are the demographics of software engineers in the world? This is a hard question that nobody has a certain answer to. Yet, without answering this question precisely, we face enormous difficulties to discuss representativeness of a sample, the needed size of the sample and, therefore, to what degree we can generalize our results. We will first introduce a way for investigating representativeness; second, discuss the issue of sample size estimation for contexts where we can estimate the size of the target population; and, third, provide a note on the ethics in sampling.

\subsection{Representativeness}
For other types of survey research, scientists often rely on demographic information published by governmental or other public bodies such as statistical offices of countries, the EU or the United Nations. These bodies are, so far, rather unhelpful for our task, because they do not provide a good idea about software-developing companies. These bodies scatter software engineering over various categories. For example, the EU's Statistical classification of economic activities in the European Community (NACE REv.~2) has different categories for software publishing, developing software for games or application hosting. More difficult, however, is that software development occurs in many other companies as well. For example, data from the German statistical office\footnote{Table 52911-0001 in \url{https://www-genesis.destatis.de}} on the usage of information and communication technologies shows that in 2018, 13~\% of \emph{all} German companies stated that they develop business information systems internally. For web-based software solutions, even 17~\% stated that they build them internally. Put together, this means that we do not have a good estimate for the number and properties of organizations that develop some kind of software.

There are possibilities to approach the demographics of software engineers in the world. There are commercial providers of data from large surveys such as Evans Data Corporation.\footnote{\url{https://evansdata.com/}} Evans Data Corporation estimated for 2018 the number of developers worldwide to be 23 million. They include information on different roles, genders, used development processes and technologies. An open alternative is the Stack Overflow Annual Developer Survey.\footnote{\url{https://insights.stackoverflow.com/survey/}} They have the bias that only people registered at Stack Overflow can be sampled. Yet, this could be tolerated in light of the popularity of the platform among software developers. They provide demographic information, for example, on whether developers are professionals, their roles, experiences and education.

With this demographic information, we can design our survey in a way that we collect comparable data as is available in the distributions for the total population. Then, we can compare the distributions in our survey and the larger surveys to estimate representativeness. This comparison should primarily be part of the interpretation and discussion of the results. This comparison prevents us from overclaiming but at the same time gives more credibility in case we cover the population well.

\subsection{Sample Size Estimation}

Having the estimate of the total number of developers world-wide, we can now ask, what would be a good size for our sample? In other contexts, we might even have better information on the population size, for example, when we want to survey GitHub developers. This is an information we can extract from GitHub itself.

There is a large body of existing work that discusses sampling techniques and suitable sample sizes. A simple way, for example, is to follow \citet{yamane1973}. He proposed to use this equation to calculate a suitable sample size $n$:

\begin{equation}
n = \frac{N}{1 + N e^2}
\end{equation}

In the equation, $N$ is the population size and $e$ is the level of precision. This level of precision is often set to 0.05 or 0.01. For the estimate of 23 million developers worldwide, how large would our sample need to be?

\begin{equation}
n = \frac{23,000,000}{1 + 23,000,000 \cdot 0.05^2} = 400
\end{equation}

So, for most intents and purposes, with a sample size of more than 400, we could claim a strong generalizability given that we also checked the representativeness as described above. Of course, most survey will fall short of this. Yet, a clear discussion comparing the sample size and representativeness with these figures makes it easy to evaluate the strength and weaknesses of a particular survey study.

For the happiness study we describe in Section~\ref{ssec:psych:example}, we assessed how happy are software developers that have GitHub accounts. We needed to contact these developers; therefore, we queried the GitHub Archive for public events providing e-mail addresses. We obtained almost $456,283$ unique e-mail addresses. We needed to find a way to sample these addresses properly. First, we conducted three pilot studies with $N=100$ randomly sampled e-mail addresses. From the studies we could estimate that roughly $98\%$ of the e-mails were delivered, and that the response rate was rather low, between $2\%$ and $4\%$. After deducting the $300$ entries from the three pilot samples, our new population size was $455,983$. With Yamane's formula, with $e = 0.05$, we found out that we required $N = 400$ complete responses. On the other hand, the formula by \citet{cochran1977}, which uses a desired value $\alpha$ for significance, suggested us to aim for $N = 664$ responses with a significance level of $\alpha = 0.01$. We opted for the more conservative value of $N = 664$ for our desired sample. That meant that we needed to send out $33,200$ e-mails assuming a $2\%$ response rate.\footnote{More information on the sampling methodology can be found in our paper~\cite{graziotin2017unhappy}.}.

\subsection{Ethics}

Sampling in survey research today almost always means soliciting answers via e-mail or social media. In a recent paper, \citet{Baltes:2016:WSI:2961111.2962628} discuss that the common practice of sending unsolicited e-mails to GitHub developers could be ethically problematic. In software engineering, there is yet no established standard or guidelines on how to conduct surveys ethically. They report that in their own surveys, the received feedback from developers on GitHub being ``spammed'' with research-related e-mails. They conclude that researchers in software engineering should discuss this issue further and create their own guidelines.

For happiness study of Section~\ref{ssec:psych:example}, as described in the previous subsection, we had to contact more than $30,000$ software developers via e-mail. Even though the developers provided a publicly listed e-mail address, we were aware that our e-mails were unsolicited and might have disturbed their daily activities. There were no available guidelines for the situation or even portals to gather volunteers for software engineering research. All we could do was to design a short and cordial invitation e-mail that, besides acting as informed consent form including ethical and privacy considerations, was of \textit{opt-in} nature. We believe that the consideration worked to a certain extent, but we also add to the experience of \citet{Baltes:2016:WSI:2961111.2962628} of receiving feedback from potential participants who were annoyed by this. While the number of complaints was not excessive, a very annoyed participant asked GitHub to check on us. After inquiring with us on the nature of the study and after inspecting our invitation e-mail, GitHub concluded that our study did not break any of their terms of services and kindly asked us to be advised before starting research activities, in the future, as they might want to check on the research design, invitation, and compliance with their terms. This last information might help future research in our field.

The \emph{Insight Association} provides ethical guidelines that consider unethical sampling, among other practices: ``Collection of respondent emails from Web sites, portals, Usenet or other bulletin board postings without specifically notifying individuals that they are being "recruited" for research purposes.''\footnote{\url{https://www.insightsassociation.org/issues-policies/best-practice/imro-guidelines-best-practices-online-sample-and-panel-management}} Hence, using GitHub or Stack Overflow information of users would not be an ethical way to contact potential survey participants.

There is no easy way out of this. We agree with Baltes and Diehl that we will probably need flexible rules and guidelines to keep developers in social media from being spammed by study requests while still allowing research to take place. In any case, we should all consider thoughtfully how and whom we contact for a survey study.

\begin{shaded}\textbf{Take Aways}

\begin{itemize}
    \item There is no suitable official data on the number and properties of software-developing companies in the world.
    \item For individual software engineers, existing demographic studies can be used to assess a survey's representativeness.
    \item For the estimate of 23 million developers worldwide, a good sample size would be 400 respondents.
    \item Ethics needs to be considered before contacting potential survey participants.
\end{itemize}

\end{shaded}

\section{Invitation and Follow-Up}

Depending on the target population, there are essentially two strategies to approach this population having both very distinct implications on the survey design and the recruitment approaches:
\begin{enumerate}
\item \emph{Closed invitations} follows the strategy of approaching known groups or individuals to participate in a survey per invitation-only and restrict the survey access only to those being invited.
\item \emph{Open invitations} follows the strategy of approaching a broader, often anonymous audience via open survey access; i.e. anyone with a link to the survey can participate.
\end{enumerate}

The first strategy allows to accurately choose the respondents based on predefined characteristics and the suitability to provide the required information, and it also allows to accurately calculate the response rate and control the participation along the data collection, e.g. by targeting those who didn't respond yet via specific requests. This increase of control by inviting subjects individually typically comes at the cost of a lower number of total responses.

The second strategy allows to spread the invitation broader, e.g. via public forums, mailing lists, social media, or at venues of conferences and workshops. This strategy is often preferred as it doesn't require to carefully select lists of subject candidates and to approach them individually, but it also comes at the cost of control in who provides the responses, thus, causing further threats that need to be carefully addressed in the survey design already. In that strategy, we need to define proper demographic questions that allow us to analyze the extent to which the respondents are eventually suitable to provide the required information (see also Sec.~\ref{sec:sampling}). 

In the NaPiRE project, for example, we started our initial runs with closed invitations. To this end, we drafted a list of subject candidates based on contacts from industry collaborations. Criteria for their inclusion were their roles and responsibilities in their respective project settings and their knowledge about not only requirements engineering, but also about how their processes were continuously improved. That is to say, we were particularly interested in surveying experienced requirements engineers which dramatically narrowed down the list of suitable candidates. The survey was then password-protected and the invitation were individualized with a clear explanation of the scope of the survey and the contained questions. When inviting our candidates, we asked them also to report to us in case they had passed the invitation to a colleague allowing us to calculate the response rates. We repeated this strategy during the first two NaPiRE runs, the second one being conducted in 10 countries in parallel and via independent invitation lists administrated individually by the respective researchers in those countries.

For the follow-up runs, we changed our instrument to focus more on current practices and problems in requirements engineering at project level taking also into account a broader spectrum of project roles (e.g., developers and architects). We further decided to open the survey and added more demographic questions that allowed us to better understand the respondents' roles and backgrounds in their projects. The distribution was then done using software engineering-related mailing lists, distribution channels of associations, such as the International Requirements Engineering Board (IREB), social media, such as Twitter, but also, again, personal contacts. We further published an IEEE Software blog post. The idea was to increase the visibility of the survey project. At the same time, we were not able to calculate the response rate and also noticed a significant drop-out rate (i.e. participants entering the survey out of curiosity and dropping out on the first survey page already). Above all, it further required an analysis of making sure that the responding population is the one of the target population and, respectively, to remove those answers from the data set clearly unrelated to the target population (e.g. respondents with no insights into the projects' requirements engineering).

Regardless of the strategy followed, it is often the case that invitees cannot participate in the survey the moment they receive an invitation despite being otherwise willing to participate. In both surveys, we therefore implemented a follow-up invitation roughly two weeks before closing the survey. To this end, we formulated, regardless of the strategy, a reminder message thanking all participants and reminding them that there is still the possibility of engaging in case they did not already.

For the happiness study (Section~\ref{ssec:psych:example}) we decided to not adopt any follow-up. Ethical reasons (see the previous section) made us decide for an opt-in mechanism. We contacted possible participants only once, at invitation time.

\begin{shaded}\textbf{Take Aways}

\begin{itemize}
    \item Both strategies to approach the target population (closed and open invitations) can be applied, but have distinct implications on the survey design and the recruitment approaches.
    \item Closed invitations are suitable in situations in which it is possible to precisely identify and approach a well-defined sample of the target population. They may also be required in situations were filtering out participants that are not part of the target population would be difficult, harming the sample representativeness.
    \item Open invitations allow reaching out for larger samples. However, they typically require more carefully considering context factors when designing the survey instruments. These context factors can then be used during the analyses to filter out participants that are not representative (e.g., applying the blocking principle to specific context factors).
\end{itemize}

\end{shaded}

\section{Alternative Approaches for Statistical Analysis}
\label{sec:statistical}

Although surveys can be qualitative (see Sec.~\ref{sec:qualitative-analysis} for more details on that analysis),
most often a majority of the questionnaires are composed of closed questions that have quantitative results. Even
for yes/no questions, we can count and calculate proportions of the answers. Therefore, and with the often
large number of participants in surveys, we usually aim at a statistical analysis of the survey results.
So, which kind of statistical analysis is reasonable for surveys?

Before we go into the different options we have for the statistical analysis, we want to discuss another important issue that is sometimes neglected: To know what we can analyze, we need to be clear what we asked for. In a survey,
we can either ask for the opinions of the participants on topics (``Automated tests are more effective than manual tests.'') or for specific facts that they experienced (``In my last project, I found more defects in the software using automated tests than manual tests.'')~\citep{torchiano2017lessons}. In the former case, we can only make an analysis of the \emph{opinion} that, for example, most people hold. Only in the latter case, we can try to analyze about facts. But even then, we need to discuss in the threats to validity that the participants' answers might be biased.

With that out of the way, we can start with the first option of statistical analysis that is always reasonable: \emph{descriptive statistics}. Afterwards, we will discuss three alternative approaches to do inference statistics which will help us to interpret the sample results for the whole population. We will cover null hypothesis significance testing, bootstrapping confidence intervals and Bayesian analysis.

\subsection{Descriptive Statistics}

The goal of descriptive statistics is to characterize the answers to one or more
questions of our specific sample. We do not yet talk about generalizing to the population.

Which descriptive statistic is suitable depends now on what we are interested in most and the scale of the data. Most often, we come across nominal, ordinal and interval scales in survey data. Nominal data are names or categories that have no order and can simply be counted. Ordinal data is what we often have in surveys where we can order the data but cannot clearly say if each point on the scale has the same distance to the next point. 

An example are the famous Likert items that range from ``I fully aggree'' to ``I fully disagree''. If we have clearly defined distances, we have interval data. Only for the latter, we can employ the full range of statistical tests.

\begin{figure}[htb]
    \centering
    \includegraphics{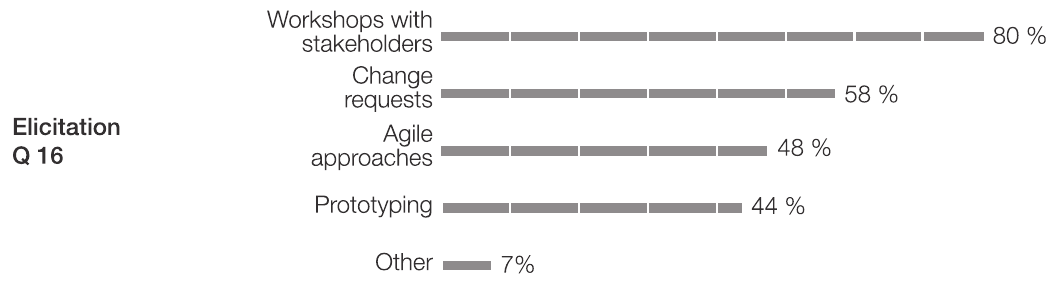}
    \caption{The proportion of respondents giving a particular answer to the question ``If you elicit requirements in your regular projects, how do you elicit them?'' visualized as bar chart \citep{MENDEZFERNANDEZ2015616}}
    \label{fig:bar-chart}
\end{figure}

For dichotomous variables, where the participants can check an answer option or not, we can calculate the proportion of the participants that checked a particular answer option. A proportion can be stated as a number between 0 and 1 or in percentages. A useful addition to giving the number is a visualization as bar chart that allows us to quickly compare many answer options. An example from the NaPiRE survey is shown in Fig.~\ref{fig:bar-chart}.

\begin{figure}[htb]
    \centering
    \includegraphics[width=.9\textwidth]{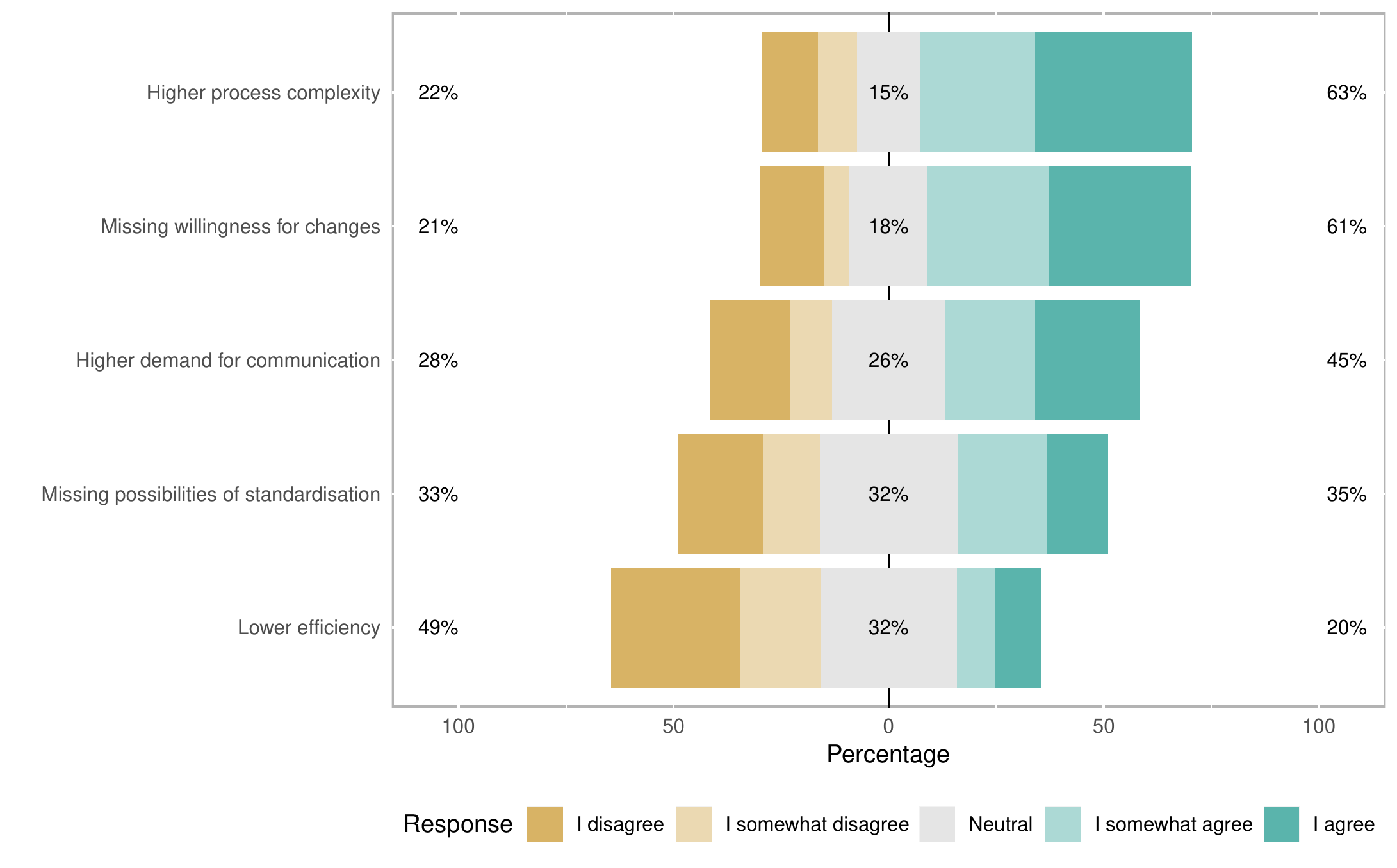}
    \caption{Stacked bar chart showing all answers to a Likert item about barriers to requirements
        engineering standards \citep{Wagner:2019:SQR:3316413.3306607}}
    \label{fig:likert}
\end{figure}

Quite common are also answers in an ordinal scale such as Likert items (``I fully agree'', ``I somewhat agree''\ldots) or frequencies (``Always'', ``Often'', ``Sometimes''\ldots). There are various descriptive statistics that we can calculate for this data. For the central tendency, we can safely use the \emph{mode}, which is simply the most
frequent answer, as well as the \emph{median}, which is the middle answer when sorting all answers~\citep{freedman2007}. To give a better understanding of the spread and dispersion of the data, we usually add the
\emph{minimum} and \emph{maximum} as well as the \emph{median absolute deviation} (MAD) or the \emph{interquartile range} (IQR). In Tab.~\ref{tab:median-et-al}, we see a table from \citep{MENDEZFERNANDEZ2015616} that used these
statistics to describe answers to Likert items about various aspects of requirements engineering. The answers
were coded from 1: I disagree to 5: I agree. In addition or alternatively, it is also quite easy to show the whole distribution of ordinal data in a stacked bar chart as shown in Fig.~\ref{fig:likert}. This particular chart
was created using the \emph{likert} package\footnote{\url{https://CRAN.R-project.org/package=likert}} in R.

\begin{table}[htb]
    \centering
    \caption{Descriptive statistics for ordinal data coded as 1 to 5 \cite{MENDEZFERNANDEZ2015616}}
    \label{tab:median-et-al}
    \begin{tabular}{p{7.5cm}rrrrr}
    \toprule
    Statement	& Mode	& Med.	& MAD	& Min.	& Max.\\
    \midrule
    The standardization of requirements engineering improves the overall process quality & 	5 &	4 &	1 &	1 &	5\\
    Offering standardized document templates and tool support benefits the communication &	5 &	4 &	1 &	1 &	5\\
    \bottomrule
    \end{tabular}
\end{table}

It is rather rare that we get data in interval scales or higher. We may get data on an interval scale when we ask for specific numbers such as the length of the last project of the participants in months. Those data can be analyzed with all available descriptive statistics such as the \emph{mean} for central tendency and \emph{variance} or \emph{standard deviation} for dispersion in addition to the ones we already have for ordinal data. A very useful visualization for such data is a boxplot, because it visualizes the distribution and allows us to identify outliers.

\subsection{Null Hypothesis Significance Testing}
\label{sec:hypothesis_testing}

Now that we have a good understanding of our sample -- and possibly are able to answer our research
questions specifically about the sample --, we want to analyze whether and what we can say about the
population we are actually interested in. This is the area of \emph{inference statistics}. To be
able to analyze something about the population, we first need hypotheses to evaluate. In our experience, unless we are conducting an exploratory study, the survey should be guided by a theory (see Sec.~\ref{sec:survey-design}).
The theory should provide propositions that can be operationalized into hypotheses to be tested with
the survey data.

The classical way to do that is to use null hypothesis significance testing (NHST). This is the usual way one is taught in statistics classes and has been used for numerous experiments and surveys. In surveys, we most often have two types of hypotheses:
\begin{itemize}
    \item Point estimate hypotheses for answers to single questions
    \item Hypotheses on correlations between answers to two questions
\end{itemize} 

For an example, we look again at \citet{MENDEZFERNANDEZ2015616}. There, we tested several hypotheses on the experience of the participants with requirements engineering in their projects. Let us look at the hypothesis H 76: \emph{Communication flaws between project team and customer are a problem.} The question in the
questionnaire was ``Considering your personal experiences, how do the following problems in requirements
engineering apply to your projects?''. The corresponding statement was ``Communication flaws between us
and the customer'' with five answer options from ``I disagree'' to ``I agree''. As above, we coded these
answers as numbers from 1 to 5. 

We operationalized the hypothesis so that the median of the data needs
to be larger than 3 (the neutral answer) so that we see the hypothesis as true. The corresponding null hypothesis is than that the median is smaller or equal to 3. 
To then test this, we employed the Wilcoxon signed rank test implemented in R. We used the rank
test as we have ordinal data which breaks the assumptions of, for example, a t-test. The result of the test is a p-value that we need to compare with our previously specified significance level (usually 0.05).
This then gives us a dichotomous answer whether we have to reject the null hypotheses and, therefore, have support for our alternative hypothesis or not. Similarly, there are statistical tests to give us p-values for which we can consider a correlation to hold true in the population.

In both cases, point estimates and correlation analysis, it is also informative and important to look at the effect sizes. A NHST only tells us whether the observed data is unlikely given the null hypothesis not how large the effect is. Especially in survey research, it is rather easy to achieve large sample sizes. The larger the sample, the more likely we get significant effects. Then the effect sizes can help us interpret the results. For correlation analysis, the correlation coefficient is already a useful effect size. For point estimates, we need an effect size that fits to the data and the used statistical test. For the Wilcoxon signed rank test above, it is often suggested to divide the test statistic $z$ by the square root of the sample size $N$ to the the standardized effect size $r$: 
\begin{equation}
    r = \frac{z}{\sqrt{N}}
\end{equation}
This can then be interpreted as 0.1 being a small effect up to 0.5 and larger as being a large effect.

There are various problems with NHST in general such as the dichotomous nature of its result~\citep{levine2008,amrhein2019}.
Yet, in our survey context, there is even one more: As discussed in Sec.~\ref{sec:sampling}, it is
in most cases very difficult to obtain a sample that is representative of the population, we want to generalize to. In such cases, it is unclear what the result of a NHST actually means. How can we generalize from non-representative data? Therefore, we need to look at alternatives.

\subsection{Alternative 1: Bootstrapping confidence intervals}

An approach that has seen considerable attention as an alternative to null-hypothesis significance testing is to use confidence intervals. The basic idea is instead of a point estimate of a p-value and a fixed threshold in the form of a significance level, we rather estimate a confidence interval around a metric we are interested in. We then rather interpret what the confidence interval means in terms of, for example, how large it is or how strongly confidence intervals of methods to compare overlap. Hence, the interpretation is not as easy as comparing the p-value with the significance level but it allows us to avoid a too simplistic dichotomous result.

To address the problem of the unclear representativeness of the sample because the population is unknown, we can further support the estimation of confidence intervals by using a resampling method. In particular, bootstrapping is helpful as it gives us asymptotically more accurate results than intervals estimated with the standard assumption of normality \citep{diciccio1996}. The idea of bootstrapping is that we repeatedly take samples with replacement and calculate the statistic we are interested in. This is repeated a large number of times and, thereby, provides us with an understanding of the distribution of the sample.

\citet{Wagner:2019:SQR:3316413.3306607} applied this approach to evaluate our theory without the use of null-hypothesis significance testing. For that, we ran 1,000 times resampling for bootstrapping confidence intervals for proportions and means of the answers to the survey questions. This works particularly well for proportions. It is problematic for questions that have answers on an ordinal scale. We discussed above that for those, we should use the median instead of the mean. As, so far, there are no established methods for bootstrapping confidence intervals for medians, we decided to work with the confidence intervals of the means but report the medians alongside of them.

In Fig.~\ref{fig:confidence-intervals}, we see the visualization of answers to a survey question as bar chart. Each bar shows the mean proportion of answers with additional black bars showing the confidence interval both derived from bootstrapping. We had a proposition for each of the answer possibilities in our theory. As we wanted to characterize what techniques are \emph{commonly} used in practice, we decided that common use should imply a proportion above 20 \%. Hence, only when the confidence interval is above 20 \%, we consider it as support for a positive proposition.

If we wanted a dichotomous decision on the propositions in the example, we would see that all confidence intervals are above 20 \% and, hence, we have support for all propositions. Yet, we can also clearly see and discuss that interviews and facilitated meetings are much more common in practice than scenarios or observation.

An alternative use of bootstrapping is for the estimation of true population means when the obtained data is not normally distributed. In the happiness study we obtained $1\,318$ questionnaire responses contributing to the SPANE-B happiness score (explained in Section~\ref{ssec:psych:example}). Our example showed strong evidence of non-normality in its distribution. Therefore, we used bootstrapping to estimate the population true mean and its confidence interval (or, how confident we are on how much developers are happy).

Bootstrapping confidence intervals has, however, also disadvantages. One problem is that it can easily be interpreted as dichotomous and would bring us back to null-hypothesis significance testing. Another problem is that it is less clear what a confidence interval means for hypotheses. When should we see support for the hypothesis, when should we not? Furthermore, there is no clear way how to integrate different sets of data, for example, from different survey runs or independent surveys.

Finally, one might argue that a disadvantage is that it is a frequentist statistical technique that interprets probabilities as relative frequencies. This brings along various assumptions \citep{kass2011statistical} that have been criticized and could, for example, be overcome by a Bayesian analysis as discussed in the following.

\subsection{Alternative 2: Bayesian analysis}
\label{sec:bayesian_analysis}

In Bayesian statistics, probability is understood as a representation of the state of knowledge or belief. It acknowledges the uncertainty in our knowledge by assigning a probability to a hypothesis instead of an accept/reject decision. Furthermore, it allows us to easily integrate existing evidence and accumulate knowledge. It does so by defining a \emph{prior} distribution. This is the distribution that describes our certainty of a hypothesis before we collected new data. 
The Bayes theorem allows us then to describe the probability of a hypothesis given the prior and new evidence. This is called the \emph{posterior} distribution.

A major difficulty with employing Bayesian data analysis instead of classical null hypothesis significance testing or classical confidence intervals is that there is not just one Bayesian technique. It is a completely different way of thinking and, thereby, there is a plethora of techniques that can resemble what we did in the frequentists methods. The most general way would be, as stated above, to calculate a probability for a hypothesis. Yet, there are many alternatives: For example, there is the concept of the \emph{Bayes factor} that can be calculated and there are standard interpretations on how strong the suppport for a hypothesis is. This is close to the way we approach the evaluation of hypothesis in NHST. Moreover, and that is the method we will describe in an example in more detail, we can also calculate confidence intervals using Bayesian methods.

For Bayesian confidence intervals, we only need three inputs: a prior distribution, data, and the level of confidence we want to have for the confidence intervals. Data we should have from the survey. The confidence level is commonly set to 0.95 but could be different if you have specific needs. The problem is usually the prior. This is one aspect of Bayesian analysis that draws a lot of criticism, because there is no mechanical way to get to it unless you have prior data. If you have prior data, for example from a previous survey, you can calculate the prior from that data. In all other cases, you have to decide on a prior. In case there is no reasonable argument for something else, the uniform distribution should be used. If another argument from theoretical considerations can be made, however, it is legitimate and useful to put in another prior. Given all inputs, there are many ways to calculate the confidence intervals. One way is, for example, to use the \emph{binom.bayes} function of the \emph{binom} R package.

When we go to the NaPiRE example, we have not yet published a Bayesian data analysis. Yet, there is data from a third run, where we apply the Bayesian confidence interval approach at the moment of writing this chapter. Here, it comes in very handy to be able to combine the data of more than one run. We do have many similarities between the second and third run. Using Bayesian analysis, we do not throw away the second run but build on it. We will look at the proposition again that workshops are commonly used in practice for eliciting requirements.

\begin{figure}[htb]
    \centering
    \includegraphics[width=.8\textwidth]{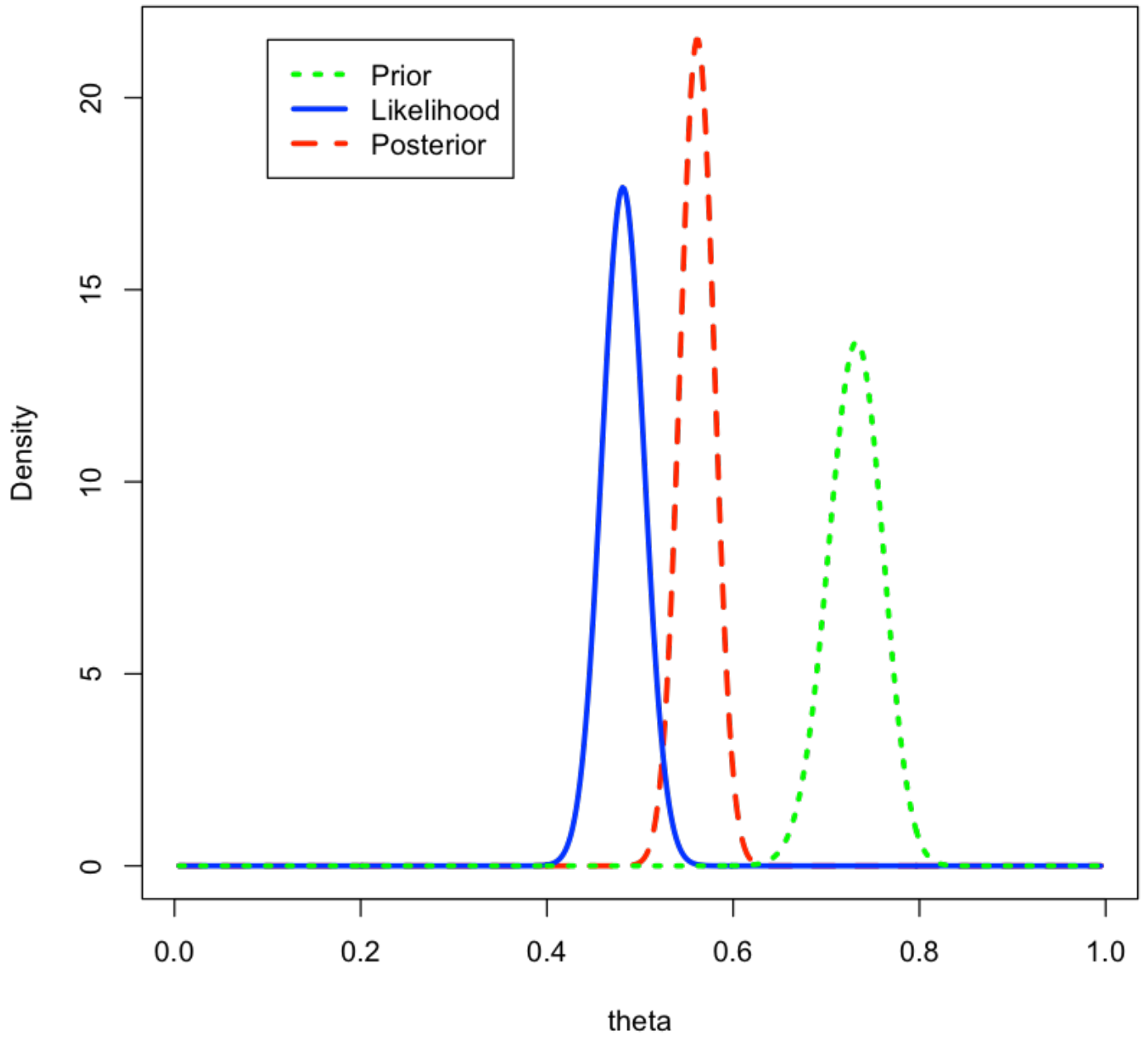}
    \caption{Graphical summary of the distribution of the used statistics in the Bayesian analysis}
    \label{fig:summary-bayes}
\end{figure}

We use the data from the second run from this question to calculate a posterior distribution given a uniform prior distribution. Commonly, beta distributions are used for that. Our R analysis gives us a posterior of beta(154, 76). Now, we can use this posterior distribution from the second run as prior for the third. Fig.~\ref{fig:summary-bayes} shows graphically how this turned out. From the second run, we had an estimation used as prior between 60~\% and 80~\% for the proportion of practitioners using workshops. The data from the third run gives us an estimate more between 40~\% and 60~\% (Likelihood). From that, we calculate the posterior that lies in between and is somewhat narrower with an estimate roughly between 50~\% and 60~\%.

With the \emph{binom.bayes} package, we can make this more precise. When we calculate 95~\% confidence intervals, it gives us a mean of 0.54 with a lower estimate of 0.51 and an upper estimate of 0.58. For interpreting this according to the hypothesis, we can again use the 20~\% threshold and confidently state that we are closer but still far away from it. We have support for the hypothesis. By using Bayesian analysis, we strengthened the evaluation of the hypothesis by including the data from two surveys and probably corrected the estimate to a range not as low as the data from the third run would suggest.

There are also disadvantages in the Bayesian approach. As for confidence intervals, we do not have a standard way of interpreting the results of the analysis. Furthermore, the tool support is not as mature as it is for the frequentist methods and, thereby, sometimes rather confusing. Furthermore, reviewers in scientific venues often know Bayesian methods less and that brings the risk of misinterpretations on their side.

% \textcolor{red}{The more general issues in using Bayesian data analysis in empirical software engineering are discussed in Chapter [11].}

\begin{shaded}\textbf{Take Aways}

\begin{itemize}
    \item Always make clear whether you aim at analyzing opinions or facts.
    \item Descriptive statistics are always helpful.
    \item Bootstrapping confidence intervals helps to deal with uncertain sampling.
    \item Bayesian analysis allows us to directly integrate prior knowledge.
\end{itemize}

\end{shaded}

\section{Qualitative Analysis}
\label{sec:qualitative-analysis}

Besides the common focus on statistical analyses, surveys can also be qualitative and contain open questions. Open questions do not impose restrictions on respondents and allow them to more precisely describe the phenomena of interest according to their perspective and perceptions. However, they can lead to a large amount of qualitative data to analyze, which is not easy and may require a significant amount of resources.

The answers to such open questions can help researchers to further understand a phenomenon eventually including causal relations among theory constructs and theoretical explanations. Hence, open questions can help generating new theories. A research method commonly employed to support such qualitative analyses is Grounded Theory \citep{glaser1992basics,strauss1990basics,charmaz2014constructing}. This method involves inductively generating theory from data \citep{glaser1967discovery}. While specific considerations for conducting and reporting grounded theory can be found from \citet{Stol2016GT}, we hereafter describe the experience and challenges of conducting the qualitative analysis in the context of the NaPiRE initiative. NaPiRE involved open questions and a large amount of data from respondents of different countries around the globe.

The first issue faced in this context was the need to translate the questionnaires into the respondents' native languages to assure that they would precisely understand the meaning of each survey question. The translations were conducted by native speakers that were part of the NaPiRE team. All translations were validated by piloting the survey with independent team members that were also native speakers. Moreover, respondents also answered in their native language. We believe that this decision allowed avoiding any confounding factor related to difficulties with the language. On the other hand, it required us a significant team coordination effort to conduct the analyses appropriately. We had to translate all the answers to English and validate the translations before starting with the analysis. The strategy that we employed was exporting all answers into a spreadsheet, creating a separate column for each answer with an automatic Google translation and then having the team validating and adjusting all translations as needed. 

The main open questions concerned causes and effects of RE problems. We asked our respondents to provide what they believe to be the main causes and effects for each of the previously selected RE problems, with one open question for the cause and another for the effect. We applied the following Grounded Theory steps on this data:

\begin{enumerate}
    \item Open coding to analyze the data by adding codes (representing key characteristics) to small coherent units in the answers, and categorizing the developed concepts in a hierarchy of categories as an abstraction of a set of codes -- all repeatedly performed until reaching a state of saturation. We define the (theoretical) saturation as the point where no new codes (or categories) are identified and the results are convincing to all participating researchers \citep{birks2011grounded}.
    \item Axial coding to define relationships between the concepts, e.g., ``causal conditions'' or ``consequences''.
    \item Internal Validation as a form of internal quality assurance of the obtained results.
\end{enumerate} 

Please note that we deviated from the Grounded Theory approach as introduced by \citet{glaser1967discovery} in two ways. First, given that we analyzed data from an anonymously conducted survey after the fact, we were not able to follow a
constant comparison approach where we iterate between the data collection and the analysis. This also means that we were not able to validate our results with
the participants, but had to rely on internal validation procedures. Second, we did not inductively build a theory from bottom up, as we started with a predefined conceptual model (i.e. the problems) whereby we did not apply selective coding to infer a central category. In our instrument, we already had a predefined set of RE problem codes for which we wanted to know how the participants
see their causes and effects. For this reason, we rely on a mix of bottom-up and
top-down approach. That is, we started with our predefined core category, namely
RE problems and a set of codes each representing one key RE problem, and sub-categories regarding the Causes and Effects, which then group the
codes emerging from the free text answers provided by the participants. We believe that similar decisions could be taken in the context of other anonymously conducted surveys relying on a predefined conceptual model. Most importantly, we highlight the importance of precisely describing the approach that has been followed and the deviations from it. 

Within the causes and effects, we again predefined the sub-categories. These sub-categories were
as follows:
\begin{itemize}
    \item For the causes, we used the sub-categories Input, Method, Organization, People, Tools suggested in our previous work on defect causal analysis \citep{kalinowski2012evidence} as we wanted to know from where in the socio-economic context the problems stem.
    \item For the implications, we use the sub-categories Customer, Design or Implementation, Product, Project or Organization, and Verification or Validation as done in our previous work \citep{fernandez2015naming} as we wanted to know where in the software project ecosystem the problems manifest themselves.
\end{itemize}     
For each answer given by the participants, we then applied open coding and
axial coding until reaching a saturation in the codes and relationships, and allocated the codes to the previously defined sub-categories. 
For coding our results, we first coded in a team of two coders the first 250
statements to get a first impression of the resulting codes, the way of formulating them and the level of abstraction for capturing the codes. After having this
overview, we organized a team of five coders within Germany and Brazil. Each
of the coders then coded approx. 200 statements for causes and additional 200
statements for effects, while getting the initial codes from the pilot phase as orientation. In case the coder was not sure how to code given statements, she marked
the code accordingly for the validation phase. During that validation phase, we
formed an additional team of three independent coders who then reviewed those
codes marked as ``needs validation'' as well as an additional sample, comprising
20\% of the statements assigned to each coder, selected on their own. After the
validation phase, we initiated a call where we discussed last open issues regarding
codes which still needed further validation, before closing the coding phase. The
overall coding process took in total three months. Despite of this huge effort, we emphasize the importance of validating qualitative analysis procedures to enhance the reliability of the results.

As a results we had information on how often a certain cause was mentioned as mechanism triggering a specific RE Problem. Similarly, on how often a consequence was mentioned as a result of an RE Problem. This allowed us to analyze the occurrence of certain RE Problem cause and effect patterns, which are reported in further detail by \citet{fernandez2017naming}.

\begin{shaded}\textbf{Take Aways}

\begin{itemize}
    \item When preparing your survey, always invest effort in avoiding confounding factors that may interfere in having respondents focusing mainly on the survey question when providing their answers (e.g., language issues). A good strategy that helps to check if this goal is properly achieved involves piloting the survey and discussing it afterwards with the pilot participants to assure that questions were easily and correctly understood. 
    \item Applying coding and analysis techniques from Grounded Theory can help to understand qualitative data gathered through open questions.
    \item When reporting the qualitative analysis of your survey, explicitly state your research method, providing details on eventual deviations.  
    \item To avoid researcher bias and improve the reliability of the results, qualitative analyses should be conducted in teams and make use of independent validations. Also, ideally the raw and analyzed data should be open to enable other researchers to replicate the analysis procedures.
\end{itemize}

\end{shaded}

\section{Issues when assessing psychological constructs}

Often, we are interested in assessing psychological constructs of survey participants. Psychological constructs are theoretical concepts to model and understand human behavior, cognition, affect, and knowledge~\citep{JohnFBinning2016}. Examples include happiness, job satisfaction, motivation, commitment, personality, intelligence, skills, and performance. These constructs can only be assessed indirectly. We cannot take out a ruler to measure the motivation of people. Yet, we need ways to proxy our measurement of a construct in robust, valid, and reliable ways.

This is why, whenever we wish to investigate psychological constructs and their variables, we need to either develop or adopt measurement instruments that are psychometrically validated. Researchers in the behavioral and social sciences refer to these validated measurements instruments as \textit{psychometrically validated psychological tests}~\citep{cohen1995}. Scientists have investigated issues of validity, reliability, bias, and fairness of psychological tests. These aspects are reflected by the word \textit{psychometrics}, which is both the act of constructing valid and reliable psychological tests as well as the branch of psychology and statistics devoted to the construction of these tests~\citep{rust2009}. Psychometrics is an established field, but software engineering has, most of the times, ignored it so far.

In this last section, we build the case for software engineering research to favor psychometric validation of tests, we introduce the very basic concepts of validity and reliability as seen by psychometric theory, which is different to how we see reliability and validity in software engineering research, and we finally describe the happiness study that we often refer as an example in the previous sections.

\subsection{Software engineering questionnaires for human participants should focus on psychometrics}

\citet{lenberg2015} have conducted a systematic literature review of behavioral software engineering studies. They found that software engineering research still has several knowledge gaps when conducting behavioral studies, and that there have been very few collaboration between software engineering and social science researchers. This missed collaboration has likely resulted in the issue that software engineering research lacks maturity when adopting or developing questionnaires to assess psychological constructs.

\citet{graziotin2015} have echoed a previous call by \citet{feldt2008} to adopt measurement instruments that come from psychology, but they argued that much research in software engineering has adopted wrong or non-validated psychological tests, and when the right test is adopted, most of the times the test items are modified towards the destruction of the test reliability and validity. Research in software engineering also fails to report on thorough evaluations of the psychometric properties of the chosen instruments. An instance of such misconduct was found by \citet{cruz2015} in their systematic literature review of 40 years of research in personality in software engineering. Although not directly mentioned by the authors, the results showed that almost half of personality studies in software engineering use the Myers-Brigg Type Indicator (MBTI), which has low to none validity and reliability properties~\citep{pittenger1993} up to being called as ``little more than an elaborate Chinese fortune cookie''~\citep{hogan2017}.

As argued by \citet{gren2018}, there is a need for a culture change in software engineering research to shift from ``seeing tool-constructing as the holy grail of research and instead value validation studies higher.'' We agree with his stance and add that the culture shift should also be from developing ad-hoc measurement instruments or tinkering with established ones to properly develop or adopt them. It is our hope, with this section, to provide motivation and background information to start this shift.

We will now provide a short overview of psychometric reliability and validity so that it becomes clearer that researchers in software engineering, when designing questionnaires that assess psychological constructs, should pay extra care when selecting tests and also when modifying existing ones. For a deeper understanding of these issues, we direct the reader to the work of \citet{gren2018}, who has offered a psychological test theory lens for characterising validity and reliability in behavioural software engineering research, our seminar works on qualitative and quantitative methodologies for behavioural software engineering, and the major textbooks and standards on this topic (e.g.,~\citet{apa2014, cohen1995, rust2009, kline2015handbook, coaley2014introduction}).

\subsection{Reliability and validity in psychometrics}

 Reliability can be seen either as the consistency of a questionnaire scores in repeated instances of it (also known as reliability/precision; for example, does a questionnaire that reveals I am extrovert tell the same if I take the same questionnaire one week from now?) or as a coefficient between scores on two equivalent forms of the same test (also known as reliability/coefficients; for example, do  two different tests on personality reveal that I am extrovert to the same or very similar degree?)~\citep{apa2014}. The reliability/coefficients can be further divided into three categories, namely alternate-form (derived by administering alternative forms of test), test-retest (derived by administering the same test on different times), and internal-consistency (derived by computing the relationship between scores derived from individual test items during a single session).

Both forms of reliability are interesting and should be kept high when developing and validating a measurement instrument. The Standards for Educational and Psychological Testing~\citep{apa2014} report that several factors influence the reliability of a measurement instrument, especially adding or removing items, changing test items, causing variations in the constructs to be measured (for example, using a test for mood to assess motivation of software developers), and administering a test to a different population than the one originally planned.

Validity in psychometrics is seen a little bit differently to what we usually mean with validity in software engineering research (see, for example, the work of Wohlin et al.~\citep{wohlin2012}). Validity in psychometrics is ``the degree to which evidence and theory support the interpretation of test scores for proposed uses of tests.''~\citep{apa2014}. What that means is that we need to ensure that any meaning we provide to the values obtained by a measurement instrument needs to be validated. \citet{rust2009} has summarised six facets of validity in the context of psychometric tests, which we now summarise. \citet{gren2018} has offered an alternative lens on validity and reliability of software engineering studies, also based on psychology, that we advise to read.

\emph{Face validity} concerns how the items of a measurement instrument are accepted by respondents. This is mostly about wording and meaning of the test questions and how they are perceived by participants. If you promise a questionnaire on software reliability but then deliver one about software testing, participants will likely feel confused or offended. Face validity is usually assessed qualitatively.

\emph{Content validity} (also known as criterion validity or domain-referenced validity) reflects how a test fits the purposes for which it was developed. If you develop a test on job satisfaction of software developers and assessed their mood instead, you have issues of content validity. Content validity is also evaluated qualitatively, because the form of deviation matters more than the degree of deviation.

\emph{Predictive validity} is assessed with the correlation between the score of a measurement instrument and a score of the degree of success in the real world. For example, a high number of years in experience in software testing is expected to have a positive correlation with ability in writing unit tests. If the correlation between these two items is higher than 0.5, the criterion for predictive validity is met and the item for years of software testing experience is retained in the test. 

\emph{Concurrent validity} is defined as the correlation of a new measurement instrument and already existing measurement instruments for the same construct. If we develop a test for assessing how developers feel motivated when at their job, we should compare the results of our test with established tests for motivation on the job. Concurrent validity assessment is very common in psychometric studies; its importance, however, is relatively secondary as the old, established measurement instruments might have low overall validity. Assessing it is, on the other hand, important for detecting issues of low validity.

\emph{Construct validity} is a major validity criterion. Constructs are not directly measurable; therefore, one way to assess how valid a measurement instrument is is to observe the relationship between the test and the phenomena that the test attempts to represent. For example, a questionnaire to assess high motivation and commitment of software developers should correlate with instances of observable high motivation and commitment of developers. It is quite difficult to assess construct validity in psychometrics, and its nature is that it is cumulative over the number of available studies.

\emph{Differential validity} assesses how scores of a test correlate with measures they should be related to, and how scores of a test do not correlate with measures they should not be related to. \citet{campbell1959} have differentiated between two aspects of differential validity, called convergent and discriminant (also called divergent) validity. Convergent validity is about correlations between constructs that are supposed to exist. A test assessing logical reasoning is supposed to correlate positively with a test assessing algorithm development. Discriminant ability is on the opposite side. A test assessing reading comprehension abilities is not supposed to strongly correlate positively with a test on algorithm development, because the two constructs are not the same. Differential validity is overall empirically demonstrated by a discrepancy between convergent validity and discriminant validity.

We have so far reported how measurement instruments implemented with questionnaires, when they are about human behaviour, cognition, affect, and knowledge face several issues of reliability and validity. Researchers in the social and behavioural sciences and statistics have spent considerable effort on developing strong methodologies and theory for implementing valid and reliable tests. Software engineering needs a cultural shift to observe and respects these issues, both when adopting and when implementing a measurement instrument. As a running example, we will now briefly summarise an experience report on adopting a psychometrically validated measurement instrument.

\subsection{An experience report on adopting a psychometrically validated instrument\label{ssec:psych:example}}

For a project on the happiness of software developers~\citep{graziotin2019,graziotin2018happens,graziotin2017unhappy} one of our goals was to estimate the distribution of happiness among software developers or, in other words, find out how happy developers are. We had a further requirement: the related questionnaire had to be as short as possible.

The first step in finding a psychometrically validated instrument to assess happiness was to find out what happiness is. We discovered two main definitions of happiness in the literature, one of which sees happiness as a sequence of experiential episodes. If we face frequent positive experiences, we are led to experiencing positive emotions and moods and appraise our existence to be a happy one. The reverse happens with negative experiences, which lead to negative affect and unhappiness. Happiness overall is the difference, or balance, between positive and negative experiences.

Once agreed on a definition, we searched for happiness and related words in academic search engines. Reading more papers led to finding new keywords and enriching our sample of candidates, which was not small at all. We then inspected all possible candidates to retain those that were short to be completed. The sample of candidates was further reduced.

We then searched all candidate names in academic search engines to look for validation studied (particular keywords here are validation, reliability, and psychometric properties). Some measurement instruments did not have any validation study beyond the one which introduced the instrument itself.

We eventually decided to adopt the Scale of Positive and Negative Experience (SPANE), which we explain in~\citet{graziotin2017unhappy}, as it is a short scale, 12 Likert items in total, on how frequently participants experience affect in the last four weeks. The introductory paper explained very carefully why and how the scale was implemented, as well as the choice of limiting the recall of experiences in the last four weeks (in short, accuracy of human memory recalling and ambiguity of people's understanding of the items themselves).

We found out that SPANE has been validated to provide high validity and reliability coefficients in (at the time) nine very large-scale psychometric studies with samples coming from different nations and cultures. Furthermore, the scale converges to other similar measurements (concurrent validity). Finally, the scale was found to be consistent across full-time workers and students. These aspects were important because the target population was sampled on GitHub, which hosts projects of developers from all around the world (indeed, we had responses from 88 different countries) and having different backgrounds and job experiences (75\% were professionals ranging from freelancing to large industries, and 15\% were students).

There was enough evidence for us to be confident in including SPANE in our studies. As we respect the hard work of those who developed and validated the scale, we included the scale verbatim in our studies. We introduced the scale using the recommended instructions, we presented the items in the same order, and we used the same Likert items as recommended.

There are several advantages of adopting a psychometrically validated scale. One of them is that we can be confident about the reliability and validity of the way we interpret the scores. In our research endeavour, we found out that software engineers have a SPANE-B (the overall SPANE score, or ``the happiness score'') centred around the values of 9-10 over a range of -24 and +24. The interpretation of this scoring is that developers are, on average, a slightly happy population. Moreover, relying on validated scales also means that often we can compare our scores with \textit{norm scores}, which are \textit{standardised scores} of several groups or populations. As many other research projects have used SPANE, we can add to our interpretation that software developers even are (just a little bit) happier than other groups of people.

A big disadvantage in adopting psychometrically validated scales lies in the complete lack of flexibility of the items, as it follows an ``all or nothing'' approach. We either include a validated scale or we do not, as changing any aspect of the items will likely invalidate the scale. SPANE has got 12 items related to emotional and affective experiences, 6 of which are positive and 6 negative. We can provide a granularity of analysis of these 12 items but nothing more than that.

\begin{shaded}\textbf{Take Aways}

\begin{itemize}
    \item Representing and assessing constructs on human behaviour, cognition, affect, and knowledge is a difficult problem that requires psychometrically validated measurement instruments.
    \item Software engineering research should either adopt or develop psychometrically validated questionnaires.
    \item Adoption or development of psychometrically validated questionnaires should consider psychometric reliability and validity issues, which are diverse and very different from the usual and common validity issues we see in ``Threats to Validity'' sections.
    \item Software engineering research should introduce studies on the development and validation of questionnaires.
\end{itemize}

\end{shaded}

\section{Recommended Further Reading}
\label{sec:further-reading}

We recommend several further book chapters and articles to complement this chapter: 
\citet{fowler2013} provides a solid general discussion on survey research in general including sampling, questions and instruments and ethics.
\citet{kitchenham2008} provided an earlier book chapter that focuses on collecting opinions by surveys but provided also more general issues relevant for survey research in software engineering. They provided more details in an older series of publications~\citep{DBLP:journals/sigsoft/PfleegerK01,DBLP:journals/sigsoft/KitchenhamP02,DBLP:journals/sigsoft/KitchenhamP02a,DBLP:journals/sigsoft/KitchenhamP02b,DBLP:journals/sigsoft/KitchenhamP02c}.
\citet{Ciolkowski2003} provide a more comprehensive process for planning and analysing a survey in software engineering.
\citet{ghazi2019survey} conducted a systematic literature review and interviews to identify common problems in software engineering surveys and also provide mitigation strategies.
For more details and methodological support on sampling, \citet{deMello2015} are a good source.
General guidelines for designing an effective survey are available from the SEI~\citep{kasunic2005designing}.
\citet{molleri2019cerse} found 39 papers with methodological aspects of surveys in software engineering that can be used as a starting point for issues not discussed (in enough depth) in this chapter.
% \textcolor{red}{Furthermore, Chapter [3] provided specific guidelines for case surveys.}

\section{Conclusion}
\label{sec:conclusion}

Survey research is becoming more and more an elementary tool in empirical software engineering as it allows to capture cross-sectional snapshots of current states of practice, i.e. they allow to describe and explain contemporary phenomena in practice (e.g. opinions, beliefs, or experiences). Survey research is indeed a powerful method and its wide adoption in the software engineering community is also steered, we believe, by the prejudice of that it is easy to employ while there exist, in fact, many non-trivial pitfalls that render survey research cumbersome. In response to this problem, the community has started to contribute hands-on experiences and lessons learnt contributions, such as by \citet{torchiano2017lessons}. 

In this chapter, we have complemented existing literature on challenges in survey research by discussing more advanced topics. Those topics range from how to use survey research to build and evaluate scientific theories over sampling and subject invitation strategies to data analysis topics considering both quantitative and qualitative data, and we complemented it with specialised use cases such as using surveys for psychometric studies. To this end, we drew from our experiences in running a globally distributed and bi-yearly replicated family of large-scale surveys in requirements engineering. While we are certainly aware of that our own journey in learning from own mistakes and slips is not done yet. We hope that by reporting and discussing these lessons we learnt over the past years, we already support other members of our research community in further improving their own survey projects.

\begin{acknowledgement}
We are grateful to all collaborating researchers in the NaPiRE initiative.
\end{acknowledgement}

\bibliographystyle{agsm}
\bibliography{survey}

\end{document}